\input harvmac
\input tables
\def\Title#1#2{\rightline{#1}\ifx\answ\bigans\nopagenumbers\pageno0\vskip1in
\else\pageno1\vskip.8in\fi \centerline{\titlefont #2}\vskip .5in}

%
\font\ticp=cmcsc10

\def\ajou#1&#2(#3){\ \sl#1\bf#2\rm(#3)}
\def\jou#1&#2(#3){\unskip, \sl#1\bf#2\rm(19#3)}

\def\frac#1#2{{#1 \over #2}}

\def\ident{1\!\!1}
\def\slash#1{#1\!\!\!\!/}
\def\ctil{\tilde c}

\lref\witten{E. Witten, 
``String Theory Dynamics in Various Dimensions,''
\ajou Nucl. Phys. &B443 (1995) 85,
hep-th/9503124.}
\lref\townsend{P. K. Townsend, 
``The eleven-dimensional supermembrane revisited,''
\ajou Phys. Lett. &350B (1995) 184,
hep-th/9501068.}
\lref\bfss{T. Banks, W. Fischler, S. Shenker, and L.Susskind, 
``M Theory as a Matrix Model: A Conjecture'',
\ajou  Phys. Rev. &D55 (1997) 5112-5128, 
hep-th/9610043.} 
\lref\mbranes{T. Banks, N. Seiberg and S. Shenker, 
``Branes from Matrices'',
\ajou Nucl.Phys. &B490 (1997) 91-106.
hep-th/9612157.}
\lref\tasi{J. Polchinski, 
``TASI Lectures on D-Branes'',
hep-th/9611050.}
\lref\acny{A. Abouelsaood, C. Callan, C. Nappi, and S. Yost,
``Open Strings in Background Gauge Fields'',
\ajou Nucl. Phys. &B280 (1987) 599.}
\lref\pair{C. Bachas and M. Porrati, 
``Pair Creation of Open Strings in an Electric Field'',
\ajou Phys. Lett. &B296 (1992) 77,
hep-th/9209032.}
\lref\bachas{C. Bachas, 
``D-Brane Dynamics'',
\ajou Phys. Lett. &B374 (1996) 37, 
hep-th/9511043.}
\lref\lifsone{G. Lifschytz, 
``Comparing D-branes to Black-branes'',
\ajou Phys. Lett. &B388 (1996) 720-726,
hep-th/9604156.}
\lref\lifstwo{G. Lifschytz, 
``Probing Bound States of D-branes'',
hep-th/9610125.}
\lref\oferberk{O. Aharony and M. Berkooz, 
``Membrane Dynamics in M(atrix) Theory'',
hep-th/9611215.}
\lref\lifsmath{G. Lifschytz and S. Mathur, 
``Supersymmetry and Membrane Interactions in M(atrix) Theory'',
hep-th/9612087.}
\lref\lifsfour{G. Lifschytz, 
``Four-brane and Six-brane Interactions in M(atrix) Theory'',
hep-th/9612223.}
\lref\vjfinn{V. Balasubramanian and F. Larsen, 
``Relativistic Brane Scattering'',
hep-th/9703039.}
\lref\cheptsey{I. Chepelev and A. Tseytlin, 
``Long-distance interactions of D-brane bound states 
and longitudinal 5-brane in M(atrix) theory'',
hep-th/9704127.}
\lref\pandp{J. Polchinski and P. Pouliot, 
``Membrane Scattering with M-momentum Transfer'',
hep-th/9704029.}
\lref\kkonbrane{G. Papadopolulos and P. K. Townsend, 
``Kaluza-Klein on the Brane'',
\ajou Phys. Lett. &B393 (1997) 59,
hep-th/9609095.}
\lref\dualbrane{E. Bergshoeff, M. de Roo, M. Green,
G. Papadopolulos and P. K. Townsend , 
``Duality of Type II 7-branes and 8-branes'',
\ajou Nucl. Phys. &B470 (1996) 113,
hep-th/9601150.}
\lref\dkps{M. Douglas, D. Kabat, P. Pouliot, and S. Shenker,
``D-branes and Short Distances in String Theory",
\ajou Nucl.Phys. &B485 (1997) 85-127,
hep-th/9608024.}
\lref\angles{M. Berkooz, M. Douglas, and R. Leigh,
``Branes Intersecting at Angles",
\ajou Nucl. Phys. &B480 (1996) 265-278,
hep-th/9606139.}
\lref\hetlife{U. Danielsson and G. Ferretti,
``The Heterotic Life of the D-particle", 
hep-th/9610082.}
\lref\power{J. Schwarz, ``The Power of M-theory",
\ajou Phys. Lett. &B367 (1996) 97-103,
hep-th/9510086}
\lref\dealwis{ S. de Alwis, ``A Note on Brane Tension and M-theory",
\ajou Phys. Lett. &B388 (1996) 291,
hep-th/9607011.}
\lref\horava{P. Horava, ``Matrix Theory and Heterotic Strings on Tori'',
hep-th/9705055.}
\lref\bdg{C. Bachas, M. Douglas, and M. B. Green,
``Anomalous Creation of Branes'',
hep-th/9705074.}
\lref\dfk{U. Danielsson, G. Ferretti and I. Klebanov,
``Creation of Fundamental Strings by Crossing D-branes'',
hep-th/9705084.}
\lref\bgl{O. Bergman, M. Gaberdiel, and G. Lifschytz,
``Branes, Orientifolds and the Creation of Elementary Strings'',
hep-th/9705130.}
\noblackbox
\Title{\vbox{\baselineskip12pt\hbox{UCSBTH-97-9}
\hbox{hep-th/9705110}
}}
{\vbox{\centerline {Interactions of Eight-branes in}
\vskip2pt\centerline{String Theory and M(atrix) Theory\ }
}}

\centerline{{\ticp John M. Pierre}}
\vskip.1in
\centerline{\sl Department of Physics}
\centerline{\sl University of California}
\centerline{\sl Santa Barbara, CA 93106-9530}
\centerline{\sl jpierre@physics.ucsb.edu}
\bigskip
\bigskip
\centerline{\bf Abstract}

We consider eight-brane configurations in M(atrix) theory and 
compute their interaction potentials with gravitons, 
membranes, and four-branes. 
We compare these results with the interactions of D8-branes with
D0-branes, D2-branes, and D4-branes in IIA string theory.
We find agreement between the two approaches for eight-brane
interactions with two-branes and four-branes.  A discrepancy
is noted in the case with zero-branes.
\Date{}


\newsec{Introduction}

In the last few years it has been discovered that the five consistent
superstring theories in ten spacetime dimensions
(IIA, IIB, Type I, $E_8 \times E_8$ Heterotic and 
$SO(32)$ Heterotic), which were 
previously thought to be distinct, are in fact different descriptions
of a single underlying theory.  Evidence for this includes various
dualities between compactified versions of these theories.
A striking realization of Witten \refs{\witten}
(which was anticipated by Townsend \refs{\townsend})
was that an eleventh dimension emerges
in the strong coupling limit of the IIA string theory.  
That is, the non-perturbative IIA
string theory is actually described by an eleven-dimensional theory (M-theory)
which has $D=11$ supergravity as its low energy effective description.

An important check of the eleven-dimensional theory is that it should be
able to describe non-perturbative objects which exist in the IIA theory,
such as the Dp-branes \refs{\tasi}.
The IIA D0,2,4,6-branes seem to have clear interpretations in 
terms of M-theory compactified on a circle
as Kaluza-Klein modes, unwrapped membranes, wrapped five-branes, and 
magnetic Kaluza-Klein branes, respectively \refs{\townsend}.
The M-theory interpretation of D8-branes remains unclear, however.
Issues related to their possible emergence from $D=11$
supergravity have been discussed in \refs{\dualbrane, \kkonbrane}.
In particular it has been suggested that the IIA 8-brane may have
an interpretation as a wrapped 9-brane, possibly related to the
``end of the world" 9-branes of M-theory with a boundary.

The remarkable conjecture of \refs{\bfss} is that the full
quantum description of M-theory in the infinite momentum
frame is given the by large $N$ limit of $D=10$ $U(N)$ supersymmetric
Yang-Mills theory dimensionally reduced to $0+1$ dimensions.
This formulation in terms of the supersymmetric quantum mechanics (SQM)
of an infinite number of highly boosted D0-branes has been 
referred to as `M(atrix) theory'.
Since M(atrix) theory does contain
eight-branes \refs{\mbranes}, we may hope to probe their
eleven-dimensional origin within this context.  A possible
relationship between D8-branes and boundary degrees of freedom
in M(atrix) theory was recently discussed in \refs{\horava}.

Scattering between various configurations of $p \leq 6$ branes in M(atrix) 
theory
has been considered in 
\refs{\oferberk, \lifsmath, \lifsfour, \vjfinn, \cheptsey, \pandp}.
Generically it seems to be the case that the low-velocity, 
long-distance potentials
between configurations of branes in M(atrix) theory
agree with those obtained from string theory and supergravity calculations.
In section two we consider configurations with M(atrix) eight-branes 
and compute
static interaction potentials with gravitons, membranes, and longitudinal
five-branes.  In section three we consider analogous configurations of
D-branes in IIA string theory, which generally contain large magnetic fields
on their world volumes.  In section four we discuss the results.  We find
agreement between the two approaches for eight-branes interacting with
two-branes or four-branes.  However in the case of eight-brane zero-brane
interactions we find that contributions coming from the $(-1)^F$R sector
in the string calculation do not appear in the M(atrix) calculation.

\newsec{8-branes in M(atrix) theory}

In this section we
consider interactions between M(atrix) eight-branes
and various lower dimensional
objects which exist in the theory using techniques developed in 
\refs{\lifsmath, \lifsfour}.  We use a gauge fixed form of the
M(atrix) theory action $(i=1,...,9)$,

\eqn\Lmatrix{L = -{1\over g}{\rm Tr}\bigg[
-{1\over 4} {\cal F}_{\mu\nu}{\cal F}^{\mu\nu} 
- {1\over 2} ({\bar D}^\mu A_\mu)^2 
+ \theta^T D_0 \theta - \theta \gamma_i [\theta,A^i]\bigg] + ghosts}
where ${\cal F}_{0i} = \partial_0 A_i - i [A_0,A_i]$,
${\cal F}_{ij} = -i[A_i,A_j]$, and all the fields are
$U(N)$ matrices.  To compute the effective action we decompose
the matrices into block diagonal backgrounds and off-diagonal fluctuations,
\eqn\fields{\eqalign{
A_i &= \pmatrix{X_i & 0 \cr 0 & x_i \cr} + \pmatrix{0 & y \cr y^T & 0 \cr}\cr
\theta &= \pmatrix{0 & \psi \cr \psi^T & 0 \cr} \cr
B &= \pmatrix{0 & b \cr b^T & 0 \cr} \cr
C &= \pmatrix{0 & c \cr c^T & 0 \cr} \cr}}
and integrate out the massive fields at one-loop.
It is possible to describe 8-branes in M(atrix) theory \refs{\mbranes} by
choosing a background in which 
$[X^1,X^2]= i c_1$, $[X^3,X^4]= i c_2$, $[X^5,X^6]= i c_3$, 
and $[X^7,X^8]= i c_4$.  This does not represent a pure 8-brane however, since
it also contains 2-branes, 4-branes, and 6-branes. 
\subsec{Interactions with gravitons}

We represent a graviton scattering off an eight-brane with the background,

\eqn\Mgrav{
\eqalign{ 
X^1 & = \pmatrix{P_1 & 0 \cr 0 & 0 \cr} \cr
X^3 & = \pmatrix{P_2 & 0 \cr 0 & 0 \cr} \cr
X^5 & = \pmatrix{P_3 & 0 \cr 0 & 0 \cr} \cr
X^7 & = \pmatrix{P_4 & 0 \cr 0 & 0 \cr} \cr
X^9 & = \pmatrix{ \ident v t & 0 \cr 0 & 0 \cr} \cr}
\qquad\eqalign{
X^2 & = \pmatrix{Q_1 & 0 \cr 0 & 0 \cr} \cr
X^4 & = \pmatrix{Q_2 & 0 \cr 0 & 0 \cr} \cr
X^6 & = \pmatrix{Q_3 & 0 \cr 0 & 0 \cr} \cr
X^8 & = \pmatrix{Q_4 & 0 \cr 0 & 0 \cr} \cr}}

\noindent
where $P_i$ and $Q_i$ are $N \times N$ matrices that 
satisfy $[P_i, Q_i] = i c_i$ in the large $N$ limit.  
The zero entry in the lower diagonal represents a single zero-brane
which has a M(atrix) theory interpretation as a graviton.  Expanding
around this background and integrating out the massive
off-diagonal modes at one-loop gives the amplitude in terms of the 
determinants (with $\tau=it$ and $\gamma=-i v$),
\eqn\gravdets{\eqalign{Bosons:& \cr
&det^{-1}(-\partial_\tau^2 + H + 2 i v)det^{-1}(-\partial_\tau^2 + H - 2 i v)\cr
&det^{-1}(-\partial_\tau^2 + H + 2 c_1)det^{-1}(-\partial_\tau^2 + H - 2 c_1)\cr
&det^{-1}(-\partial_\tau^2 + H + 2 c_2)det^{-1}(-\partial_\tau^2 + H - 2 c_2)\cr
&det^{-1}(-\partial_\tau^2 + H + 2 c_3)det^{-1}(-\partial_\tau^2 + H - 2 c_3)\cr
&det^{-1}(-\partial_\tau^2 + H + 2 c_4)det^{-1}(-\partial_\tau^2 + H - 2 c_4)\cr
Ghosts:&\cr
&det^{2}(-\partial_\tau^2 + H)\cr
Fermions:&\cr
&det(\partial_\tau + \slash{m}_f)\cr}}
where,
$$H=P_1^2 + Q_1^2 + P_2^2 + Q_2^2 + P_3^2 + Q_3^2 + P_4^2 + Q_4^2 + \ident v^2 t^2$$
and,
$$\slash{m}_f=\gamma_1 P_1 + \gamma_2 Q_1 + \gamma_3 P_2 + \gamma_4 Q_2
+ \gamma_5 P_3 + \gamma_6 Q_3 + \gamma_7 P_4 + \gamma_8 Q_4
+ \gamma_9 v t.$$

\noindent
The $P_i^2 + Q_i^2$ terms in $H$ are just a collection of simple harmonic oscillator
Hamiltonians with eigenvalues $c_i(2 n_i +1)$. Similarly 
$-\partial_\tau^2 + \gamma^2\tau^2$ has eigenvalues $2\gamma(n+{1\over 2})$.
We can easily compute the bosonic and ghost determinants, and using Schwinger's
proper time representation, they give a contribution to the one-loop
effective action of
\eqn\gravB{\eqalign{\Gamma_B &= i \int_0^\infty {ds\over s}{1\over 2\sin vs}
{1\over 16\sinh c_1s\sinh c_2s\sinh c_3s\sinh c_4s}\cr
&\qquad\times\big\{2 \cosh 2 v s  - 2\cr
&\qquad + 2\cosh 2 c_1 s + 2\cosh 2 c_2 s+ 2\cosh 2 c_3 s + 2\cosh 2 c_4 s
\big\}\cr}}

The fermionic determinant is problematic because it cannot be converted into
Klein-Gordon form by the usual method (there is no tenth $16\times 16$ matrix which
anticommutes with all the $\gamma_i$'s).  We can however evaluate it in the 
$v\rightarrow 0$ limit and the result is,
\eqn\gravF{\eqalign{\Gamma_F &= i \int_0^\infty {ds\over s}{1\over 2 v s}
{1\over 16\sinh c_1s\sinh c_2s\sinh c_3s\sinh c_4s}\cr
&\qquad\times\bigg\{ 
-8\cosh c_1 s\cosh c_2 s \cosh c_3 s\cosh c_4 s + {\cal O}(v) \bigg\}\cr}}
From \gravB and \gravF we extract the effective potential which is defined
through 
\eqn\Vdef{\Gamma = \Gamma_B + \Gamma_F = -i \int d\tau V(R^2=v^2\tau^2).}  By
expanding the integrand for small $c_i$ we can do the integral over $s$
by analytic continuation and we obtain the result,

\eqn\Vgrav{V = \bigg\{{c_1^4 + c_2^4 + c_3^4 +  c_4^4 
-2(c_1^2 c_2^2 + c_1^2 c_3^2 + c_1^2 c_4^2 + c_2^2 c_3^2 + c_2^2 c_4^2 + c_3^2 c_4^2)
\over 16 c_1 c_2 c_3 c_4} + {\cal O}(v)\bigg\}R}

\subsec{Interactions with membranes}

We represent a membrane scattering off an eight-brane with the background,

\eqn\Mmem{
\eqalign{ 
X^1 & = \pmatrix{P_1 & 0 \cr 0 & p_1 \cr} \cr
X^3 & = \pmatrix{P_2 & 0 \cr 0 & 0 \cr} \cr
X^5 & = \pmatrix{P_3 & 0 \cr 0 & 0 \cr} \cr
X^7 & = \pmatrix{P_4 & 0 \cr 0 & 0 \cr} \cr
X^9 & = \pmatrix{ \ident v t & 0 \cr 0 & 0 \cr} \cr}
\qquad\eqalign{
X^2 & = \pmatrix{Q_1 & 0 \cr 0 & q_1\cr} \cr
X^4 & = \pmatrix{Q_2 & 0 \cr 0 & 0 \cr} \cr
X^6 & = \pmatrix{Q_3 & 0 \cr 0 & 0 \cr} \cr
X^8 & = \pmatrix{Q_4 & 0 \cr 0 & 0 \cr} \cr}}

\noindent
where $P_i$ and $Q_i$ are as before, while $p_1$ and $q_1$ are $n \times n$
matrices which we take to satisfy $[p_1, q_1] = i c$.  In
this case we also take $c_1=c_2=c_3=c_4=c$. The
determinants are,
\eqn\membdets{\eqalign{Bosons:& \cr
&det^{-2}(-\partial_\tau^2 + H) \cr
&det^{-3}(-\partial_\tau^2 + H + 2 c)det^{-3}(-\partial_\tau^2 + H - 2 c) \cr
&det^{-1}(-\partial_\tau^2 + H + 2 i v)det^{-1}(-\partial_\tau^2 + H - 2 i v) \cr
Ghosts:&\cr
&det^{2}(-\partial_\tau^2 + H)\cr
Fermions:&\cr
&det(\partial_\tau + \slash{m}_f)\cr}}
where,
$$H=(P_1+p_1)^2 + (Q_1-q_1)^2 
+ P_2^2 + Q_2^2 + P_3^2 + Q_3^2 + P_4^2 + Q_4^2 + \ident v^2 t^2$$
and,
$$\slash{m}_f=\gamma_1 (P_1+p_1) + \gamma_2 (Q_1-q_1) + \gamma_3 P_2 + \gamma_4 Q_2
+ \gamma_5 P_3 + \gamma_6 Q_3 + \gamma_7 P_4 + \gamma_8 Q_4
+ \gamma_9 v t.$$

\noindent
Defining $P_{\pm}= P_1 \pm p_1$ and $Q_{\pm} = Q_1 \pm q_1$ which satisfy
$[P_+,Q_+]=[P_-,Q_-]=2 i c$ and $[P_+,Q_-]=[P_-,Q_+]=0$,we can represent
$Q_\pm$ by $x_\pm$ and $P_\pm$ by $2i c \partial_{x\pm}$
so that $H$ has eigenvalues,
$$H=4 c^2 k^2_+ + x_-^2 + c (2 n_2 +2 n_3 + 2 n_4 + 3) + \gamma^2\tau^2$$
where $k_+$ and $x_-$ are continuous.
The bosonic and ghost determinants give a contribution of,
\eqn\memB{\eqalign{\Gamma_B &= 
i {\cal N}_1 \int_0^\infty {ds\over s} {1\over 4 c s}
{1\over 2\sin vs} {1\over 8\sinh^3 c s}
\big\{2 \cos 2vs + 6 \cosh 2 cs\big\}}}
where the overall factor of ${\cal N}_1 = \int dx_+ {dk_- \over 2\pi}$ comes from
the degeneracy in $H$.  As before, we compute the fermionic determinant
in the $v\rightarrow 0$  limit and it gives a contribution of,
\eqn\memF{\eqalign{\Gamma_F &= 
i {\cal N}_1 \int_0^\infty {ds\over s} {1\over 4 c s}
{1\over 2ivs} {1\over 8\sinh^3 c s}
\big\{-8\cosh^3cs + {\cal O}(v)\big\}}}
The potential extracted from \memB and \memF is,

\eqn\Vmem{V = {\cal N}_1 \big\{{3 \over 32} + {\cal O}(v)\big\} R}
 
\subsec{Interactions with longitudinal five-branes}

We represent a four-brane scattering off an eight-brane with the background,

\eqn\Mfour{
\eqalign{ 
X^1 & = \pmatrix{P_1 & 0 \cr  0 & p_1\cr} \cr
X^3 & = \pmatrix{P_2 & 0 \cr 0 & p_2 \cr} \cr
X^5 & = \pmatrix{P_3 & 0 \cr 0 & 0 \cr} \cr
X^7 & = \pmatrix{P_4 & 0 \cr 0 & 0 \cr} \cr
X^9 & = \pmatrix{ \ident v t & 0 \cr 0 & 0 \cr} .\cr}
\qquad\eqalign{
X^2 & = \pmatrix{Q_1 & 0 \cr 0 & q_1  \cr} \cr
X^4 & = \pmatrix{Q_2 & 0 \cr 0 & q_2  \cr} \cr
X^6 & = \pmatrix{Q_3 & 0 \cr 0 & 0 \cr} \cr
X^8 & = \pmatrix{Q_4 & 0 \cr 0 & 0 \cr} \cr}}

\noindent
In this case $p_i$ and $q_i$ are $n \times n$
matrices which we take to satisfy $[p_1, q_1] = i c_1$, $[p_2, q_2] = i c_2$.
The four-brane configuration in the lower block of the $X_i$'s carries membrane as
well as longitudinal five-brane charge \refs{\mbranes}.  
The one-loop amplitude is given in terms of
the determinants, 
\eqn\fourdets{\eqalign{Bosons:& \cr
&det^{-4}(-\partial_\tau^2 + H)\cr
&det^{-1}(-\partial_\tau^2 + H + 2 i v)det^{-1}(-\partial_\tau^2 + H - 2 i v)\cr
&det^{-1}(-\partial_\tau^2 + H + 2 c_3)det^{-1}(-\partial_\tau^2 + H - 2 c_3)\cr
&det^{-1}(-\partial_\tau^2 + H + 2 c_4)det^{-1}(-\partial_\tau^2 + H - 2 c_4)\cr
Ghosts:&\cr
&det^{2}(-\partial_\tau^2 + H)\cr
Fermions:&\cr
&det(\partial_\tau + \slash{m}_f)\cr}}
where,
$$H=(P_1+p_1^2) + (Q_1-q_1)^2 + (P_2+p_2)^2 + (Q_2-q_2)^2 
+ P_3^2 + Q_3^2 + P_4^2 + Q_4^2 + \ident v^2 t^2$$
and,
$$\slash{m}_f=\gamma_1 (P_1+p_1) + \gamma_2 (Q_1-q_1) 
+ \gamma_3 (P_2+p_2) + \gamma_4 (Q_2-q_2)
+ \gamma_5 P_3 + \gamma_6 Q_3 + \gamma_7 P_4 + \gamma_8 Q_4
+ \gamma_9 v t.$$
The spectrum of $H$ is,
$$H=4 c_1^2 k_{1+}^2 + 4 c_2^2 k_{2+}^2 + x_{1-}^2 + x_{2-}^2
+ c_3(2 n_3 + 1) + c_4(2 n_4 + 1) + \gamma^2 \tau^2.$$
and the  contribution to the one-loop integral from bosons and ghosts is,
\eqn\fiveB{\eqalign{\Gamma_B &= {\cal N}_1 {\cal N}_2 \int_0^\infty {ds \over s} 
{1\over c_1 c_2} {1\over (4 s)^2}
{1\over 2 i\sin v s}
{1\over 4\sinh c_3 s\sinh c_4 s} \cr
&\qquad\times\big\{2 + 2\cos 2 v s +2\cosh 2 c_3 s + 2\cosh 2 c_4 s\big\}\cr}}
where again there are degeneracies coming from 
${\cal N}_i = \int dx_{i+}{dk_{i-}\over 2\pi}$.
The fermionic contribution in the $v\rightarrow 0$ limit is
\eqn\fiveF{\eqalign{\Gamma_F &= {\cal N}_1 {\cal N}_2 \int_0^\infty {ds \over s} 
{1\over c_1 c_2} {1\over (4 s)^2}
{1\over 2 i v s}
{1\over 4\sinh c_3 s\sinh c_4 s}\{-6\cosh c_3 s \cosh c_4 s + {\cal O}(v) \}\cr}}
The potential extracted from \fiveF and \fiveB is
\eqn\Vfive{V = {{\cal N}_1 {\cal N}_2 \over c_1 c_2} {1 \over 64\pi}
\bigg\{ {(c_3^2 - c_4^2)^2 \over c_3 c_4} + {\cal O}(v) \bigg\} R}

\newsec{D8-brane interactions in IIA string theory}

In this section we compute interactions involving D8-branes in Type IIA string
theory.  To reproduce the brane configurations of the M(atrix) theory we
must turn on a large constant magnetic field in the world volumes of the
 D-branes.  This has the effect of forming a non-marginal
bound state with lower dimensional branes.  Using the techniques of 
\refs{\tasi,\pair,\bachas,\lifsone,\lifstwo} we compute the one-loop vacuum
amplitude for open strings stretched between these D-brane configurations,
including the effects of the relative motions and background magnetic fields.
As discussed in \refs{\dkps} the short distance limit of these amplitudes
is dominated by the lightest modes of the open strings ending on the branes,
and this is the physics which is encoded in the SQM of the M(atrix) theory zero-branes.
In the long distance limit the amplitudes are dominated by the exchange
of massless closed strings.
What is somewhat surprising is that the long and short distance potentials
agree, even though the configurations do not in general preserve any
supersymmetry. This can be explained by the fact that we must consider
very strong magnetic fields, which correspond to a large number of
zero-branes bound to the system.  The zero-branes then dominate the dynamics
and the configuration becomes ``almost" supersymmetric \refs{\lifsmath}. 
The M(atrix) theory analog of this is that the zero-branes are boosted to
the infinite momentum frame in the large $N$ limit. 

\subsec{Pure D8-branes}

Let us first review the interactions of D8-branes in the absence of a 
background magnetic field \refs{\lifsone}.  The one-loop vacuum amplitude
gives the scattering phase shift for moving D-branes  \refs{\bachas} with
relative velocity $v=\tanh\pi\nu$.
For Dp-branes ($p=4,6$) scattering off D8-branes the amplitude is
\foot{Our conventions for the $f$-functions are as in \refs{\tasi}.
The $\Theta$-functions are standard.},
\eqn\Dpure{\eqalign{{\cal A} &= {1\over 2\pi} 
\int {dt\over t} (4\pi t)^{-p/2}Z_B \times Z_F \cr
Z_B &= {\Theta'_1(0|it)\over \Theta_1(\nu t|i t)} f_1^{-p} f_4^{p-8} \cr
Z_F &= \bigg\{{\Theta_3(\nu t|it) \over \Theta_3(0|i t)} f_3^p f_2^{8-p}
           - {\Theta_2(\nu t|it) \over \Theta_2(0|i t)} f_2^p f_3^{8-p}\bigg\}\cr}}
The one-loop vacuum amplitude for a static D0-brane and D8-brane is given by,
\eqn\Dzero{\eqalign{{\cal A} &= \int {dt\over t} (4\pi t)^{1/2} f_4^{-8}
{1\over 2}\big\{f_2^8 - f_3^8 + f_4^8\big\}\cr}}
The three terms in brackets come from the NS, R, and $(-1)^F$R sectors respectively,
and the expression vanishes by the `abstruse identity' reflecting the fact that this
configuration preserves a quarter of the supersymmetries.  This is a case where
the number of directions with mixed Neumann-Dirichlet boundary conditions is eight
and therefore the oscillator expansions for transverse bosons and R fermions 
have half-integer moding, while transverse NS fermions have integer moding.
The transverse NS fermions then have zero modes which is why the $(-1)^F$ NS does not
contribute to the partition function.  We can in principle compute velocity
dependent corrections and they appear as in \Dpure, except for the $(-1)^F$R
sector which has an additional complication due to the superghost zero modes.
We hope to address this case in more detail in future work.

\subsec{The 8-6-4-2-0 configuration}

We turn on a constant background world volume field strength,

\eqn\Fuv{ F^{\mu\nu} =
\pmatrix{ 0 & 0 & 0 & 0 & 0 & 0 & 0 & 0 & 0\cr
          0 & 0 &F_1& 0 & 0 & 0 & 0 & 0 & 0\cr
          0 &-F_1& 0 & 0 & 0 & 0 & 0 & 0 & 0\cr
          0 & 0  & 0 & 0 &F_2& 0 & 0 & 0 & 0\cr
          0 & 0  & 0 &-F_2& 0 & 0 & 0 & 0 & 0\cr
          0 & 0 & 0 & 0 & 0 & 0 &F_3 & 0 & 0\cr
          0 & 0 & 0 & 0 & 0 &-F_3 & 0 & 0 & 0\cr
          0 & 0 & 0 & 0 & 0 & 0 & 0 & 0 & F_4\cr
          0 & 0 & 0 & 0 & 0 & 0 & 0 &-F_4 & 0\cr}}

\noindent
so that the D8-brane carries 6,4,2, and 0 brane charge
due to couplings with the other R-R potential forms through,

$$C_7 \wedge F + {1 \over 2} C_5 \wedge F \wedge F
+ {1 \over 3!} C_3 \wedge F \wedge F \wedge F
+ {1 \over 4!} C_1 \wedge F \wedge F \wedge F \wedge F.$$
This describes a non-marginal bound state of 8-6-4-2-0 branes analogous
to the M(atrix) eight-brane configuration.  Let us consider interactions
of D0-branes with this object.  The magnetic field alters the boundary conditions at
 the open string endpoints \refs{\acny}
 on the D8-brane
(which we take to be at $\sigma=0$),

\eqn\FDbc{\eqalign{
\partial_\sigma X^i + F^i_\nu \partial_\tau X^\nu &=0   \cr
\partial_\tau X^i & =0 \cr
\partial_\tau X^9 & =0 \cr}
\qquad\eqalign{\sigma &=0 \cr \sigma &=\pi \cr \sigma &=0,\pi \cr}}
In contrast to the case with a pure D8-brane, this configuration
preserves no supersymmetries.
The Dirichlet boundary conditions at the endpoint 
on the D0-brane ($\sigma=\pi$)
flip the moding of the NS and R sectors to integral
and half-integral, respectively.  However, the constant magnetic field also alters
the boundary conditions and mode expansions of the transverse fermionic fields 
so that there are no zero modes, therefore
all sectors contribute to the partition function.  The one-loop vacuum
amplitude for open strings stretched between this configuration is,

\eqn\IIAzero{\eqalign{{\cal A} &= {1\over 2\pi}\int {dt\over t} Z_B \times Z_F \cr
Z_B &= {\Theta'_1(0|i t)\over \Theta_1(\nu t|i t)}
\Theta_4(i\epsilon_1 t|i t)^{-1}
\Theta_4(i\epsilon_2 t|i t)^{-1}
\Theta_4(i\epsilon_3 t|i t)^{-1}
\Theta_4(i\epsilon_4 t|i t)^{-1} \cr
Z_F &= {1\over 2}\bigg\{
\Theta_2(i\epsilon_1 t|i t)\Theta_2(i\epsilon_2 t|i t)
\Theta_2(i\epsilon_3 t|i t)\Theta_2(i\epsilon_4 t|i t) \cr
&\qquad - \Theta_1(i\epsilon_1 t|i t)\Theta_1(i\epsilon_2 t|i t)
\Theta_1(i\epsilon_3 t|i t)\Theta_1(i\epsilon_4 t|i t) \cr
&\qquad-\Theta_3(i\epsilon_1 t|i t)\Theta_3(i\epsilon_2 t|i t)
\Theta_3(i\epsilon_3 t|i t)\Theta_3(i\epsilon_4 t|i t) \cr
&\qquad + \Theta_4(i\epsilon_1 t|i t)\Theta_4(i\epsilon_2 t|i t)
\Theta_4(i\epsilon_3 t|i t)\Theta_4(i\epsilon_4 t|i t) + {\cal O}(\nu)\bigg\}\cr}}
where the four terms in brackets come from the NS, $(-1)^F$NS,
R, and $(-1)^F$R sectors, respectively and $\tan \pi \epsilon_i = F_i$.
Here we are considering an adiabatic approximation where the D-branes are 
moving with a vanishingly small relative velocity, so we only will keep the
leading ($v^{-1}$) term in \IIAzero.
Since we want to consider large values of $F_i$ we make a change of variables
to $\epsilon_i = {1\over 2} - \ctil_i$ and take $\ctil_i$ to be small.
We can truncate the amplitude to include only the lightest open string modes
by taking the $t\rightarrow\infty$ limit of \IIAzero and we find,
\eqn\openzero{\eqalign{{\cal A} &\rightarrow {1\over 2\pi}\int {dt\over t}
{1 \over \nu t}
{1\over 16\sinh\pi\ctil_1 t\sinh\pi\ctil_2 t\sinh\pi\ctil_3 t\sinh\pi\ctil_4 t}\cr
&\quad\times\bigg\{ 2\cosh 2\pi\ctil_1 t + 2\cosh 2\pi\ctil_2 t +
2\cosh 2\pi\ctil_3 t + 2\cosh 2\pi\ctil_4 t \cr
&\qquad -8 \cosh\pi\ctil_1 t\cosh\pi\ctil_2 t\cosh\pi\ctil_3 t\cosh\pi\ctil_4 t\cr
&\qquad +8 \sinh\pi\ctil_1 t\sinh\pi\ctil_2 t\sinh\pi\ctil_3 t\sinh\pi\ctil_4 t
+ {\cal O}(\nu)\bigg\}\cr}}
This is the short distance limit, where we would expect that the amplitude
should agree with the M(atrix) results $\Gamma \sim i{\cal A}$ \gravB\gravF.
However the term coming from the $(-1)^F$R sector in \openzero does not seem to be
accounted for in the M(atrix) calculation; otherwise we find agreement when
$c_i = \pi\ctil_i$ and $v \approx \pi\nu$.

To find the long distance potential defined through 
${\cal A} = - \int d\tau V(R^2=v^2\tau^2)$
we take the $t\rightarrow 0$ limit of \IIAzero which is dominated by the
exchange of light closed strings between the D-branes.  By expanding the
integrand to lowest order in $c_i$ and doing the integral over $t$ by analytic
continuation we arrive at the following result,

\eqn\Vzero{V = - {\Gamma({-1/ 2})\over 32 \sqrt{\pi}}
{c_1^4 + c_2^4 + c_3^4 + c_4^4 
- 2(c_1^2 c_2^2 + c_1^2 c_3^2 + c_1^2 c_4^2 + c_2^2 c_3^2 + c_2^2 c_4^2 + c_3^2 c_4^2)
+ 8 c_1 c_2 c_3 c_4 \over c_1 c_2 c_3 c_4} R}
This static linear potential vanishes when $c_1=c_2=c_3=c_4$ 
This can be understood by a T-duality which
takes this system to a pair of non-intersecting D4-branes where we
know that some supersymmetry will be preserved when they are oriented at
$SU(4)$ angles \refs{\angles}.  The M(atrix) result \Vgrav, which is missing the last
term in \Vzero, does not exhibit this important behavior.

\subsec{2-0 and 8-6-4-2-0 interactions}

In this case we must also turn on a magnetic field in the
world volume of the 2-brane, which we take to be $F^{12}=-F^{21}=F$.
For simplicity we also set $F_1=F_2=F_3=F_4=F$ in \Fuv.  In
the directions common to the 2-brane and 8-brane, the boundary conditions are
the same as for an open string with opposite charges on the
ends in a constant magnetic field , 
so we get an overall factor of $(1+F^2)$ in the partition function 
\refs{\acny}.
The one-loop vacuum amplitude for open strings stretched between
this configuration is given by,

\eqn\IIAtwo{\eqalign{{\cal A} &= L^2 {1 + F^2 \over 2\pi} 
{\int} {dt \over t} (4 \pi t)^{-1} Z_B \times Z_F \cr
Z_B &= {\Theta'_1(0|it) \over \Theta_1(\nu t|it)} f_1^{-2} 
\Theta_4(i\epsilon t|i t)^{-3} \cr
Z_F &= {1 \over{2}} \bigg\{
{\Theta_3(\nu t|it)\over{\Theta_3(0|it)}} f_3^2 \Theta_2(i \epsilon t|it)^3
- {\Theta_2(\nu t|it)\over{\Theta_2(0|it)}} f_2^2 \Theta_3(i \epsilon t|it)^3 \cr 
&\qquad - i {\Theta_4(\nu t|it)\over{\Theta_4(0|it)}} f_4^2 \Theta_1(i \epsilon t|it)^3
\bigg\}\cr}}

\noindent
where the three terms in $Z_F$ come from the NS, R and $(-1)^F$ NS sectors
respectively and $\tan \pi \epsilon = F$.  By T-duality, $Z_B$ and $Z_F$ 
are the same as for the 0-brane, 6-brane configuration considered in
\refs{\lifsfour}. 
By expanding around a large
magnetic field value $\epsilon = {1\over 2} - \ctil$ and going to the limit
$t \rightarrow \infty$ where massless open string
modes dominate the amplitude becomes,

\eqn\opentwo{\eqalign{{\cal A} &\rightarrow L^2 {1+F^2\over 2\pi} \int {dt \over t}
{1\over 4 \pi t} {\pi \over 8\sin \pi\nu t \sinh^3 \pi\ctil t} \cr
&\qquad\times\bigg\{ 2 \cos 2\pi\nu t + 6\cosh 2\pi\ctil t
-8\cos\pi\nu t \cosh^3\pi\ctil t \bigg\}.}}

The long distance behavior coming from the massless closed string 
channel is obtained from \IIAtwo
in the $t\rightarrow 0$ limit.  The amplitude is divergent,
but we can regularize the situation by extracting an effective potential
${\cal A} = - \int_{-\infty}^{+\infty}d\tau V(v^2\tau^2)$.  After
relabeling $\pi\nu=v$, $\pi\ctil=c$, $v\tau=R$; expanding to lowest
order in $c$ and $v$; and doing the integral over $t$ by analytic continuation
we get the result,

\eqn\Vtwo{V = L^2 {(1+F^2)\over 32\pi}{(v^4 + 6c^2v^2-3c^4) \over c^3} R.}  
In the limit where $F\approx {1\over c}$ is very large and
$v\rightarrow 0$ we find precise
agreement between \opentwo,\Vtwo and the M(atrix) results \gravB,\gravF,
and \Vgrav provided that we identify ${\cal N}_1 = {L^2 \over \pi c}$ as in
\refs{\lifsmath,\lifsfour}.

\subsec{4-2-0 and 8-6-4-2-0 interactions}

To reproduce the analogous M(atrix) configuration we must turn on a 
constant magnetic field in the D4-brane world volume
$F^{12}=-F^{21}=F_1$, $F^{34}=-F^{43}=F_2$. 
The one-loop vacuum amplitude for open strings stretched between
this configuration is given by,

\eqn\IIAfour{\eqalign{{\cal A} &= L^4 {(1 + F_1^2)(1 + F_2^2) \over 2\pi} 
{\int} {dt \over t} (4 \pi t)^{-2} Z_B \times Z_F \cr
Z_B &= {\Theta'_1(0|it) \over \Theta_1(\nu t|it)} f_1^{-4} 
\Theta_4(i\epsilon_3 t|i t)^{-1} \Theta_4(i\epsilon_4 t|i t)^{-1} \cr
Z_F &= {1 \over{2}} \bigg\{
{\Theta_3(\nu t|it)\over{\Theta_3(0|it)}} f_3^4 
\Theta_2(i \epsilon_3 t|it)\Theta_2(i \epsilon_4 t|it) \cr
&\qquad - {\Theta_2(\nu t|it)\over{\Theta_2(0|it)}} f_2^4 
\Theta_3(i \epsilon_3 t|it) \Theta_3(i \epsilon_4 t|it) \cr 
&\qquad + {\Theta_4(\nu t|it)\over{\Theta_4(0|it)}} f_4^4 
\Theta_1(i \epsilon_3 t|it)\Theta_1(i \epsilon_4 t|it)
\bigg\}\cr}}
where $\tan\pi\epsilon_i = F_i$.
In the $t\rightarrow \infty$ limit the amplitude becomes,

\eqn\openfour{\eqalign{{\cal A} &\rightarrow L^4 {(1+F_1^2)(1+F_2^2)\over 2\pi}
\int{dt\over t} (4\pi t)^{-2} 
{\pi\over 4\sin\pi\nu t \sinh\pi\ctil_1 t \sinh\pi\ctil_2 t} \cr
&\qquad\times\bigg\{2+2\cos 2\pi\nu t + 2\cosh 2\pi\ctil_1 t +2\cosh 2\pi\ctil_2 t
-8 \cos\pi\nu t \cosh\pi\ctil_1 t \cosh\pi\ctil_2 t\bigg\}}}
with $\epsilon_i = {1\over 2} - \ctil_i$.

In the $t\rightarrow 0$ limit we extract the long distance potential
to lowest order in $c_i = \pi\ctil_i$ and $v = \pi\nu$ and find,

\eqn\Vfour{V = L^4 {(1+F_1^2)(1+F_2)^2 \over 64\pi^2}
{v^4 + 2(c_1^2 + c_2^2) v^2 + (c_1^2 - c_2^2)^2 \over c_1 c_2} R.}
When $c_1=c_2$ the potential vanishes at zero velocity, reflecting the
fact that some supersymmetry is preserved. 
This is related by T-duality to D2-branes at $SU(2)$ angles\refs{\angles}.
In the large $F_i \approx {1\over c_i}$ and $v\rightarrow 0$ limit there is
agreement between \openfour,\Vfour and the M(atrix) results
\fiveB, \fiveF, \Vfive when we identify ${\cal N}_i = {L^2 \over \pi c_i}$

\newsec{Discussion}

We have computed potentials between various configurations containing 8-branes
in M(atrix) theory and IIA string theory and compared their velocity
independent terms.  For the cases of 8-branes interacting with 2-branes
and 4-branes we find agreement between the two approaches
\foot{We also checked interactions between two 8-branes and found 
agreement between string theory and M(atrix) results.  There is
no static potential and velocity dependent corrections appear at
order $v^4$.  Velocity dependent corrections to the potentials for
8-branes interacting with 2-branes and 4-branes appear to agree
in the two approaches as well.}.  However in
the case with 8-branes and 0-branes we find that the M(atrix) approach
fails to include contributions coming from the $(-1)^F$ R sector in the
string calculation.  It may be the case that extra degrees of freedom need
to be added to the M(atrix) model, and we note that
the addition of a bosonic determinant such as 
$det^{-1/2}(-\partial_\tau^2 + \gamma^2\tau^2)$ to \gravdets
would lead to agreement.  It is possible that
the calculation itself breaks down, since for example
we cannot rely on large values of an impact parameter to control instabilities
and as noted in \refs{\hetlife} we are somewhat outside of the region of validity 
of the eikonal approximation in this case.  The agreement obtained for
2-branes and 4-branes would however seem to suggest that the technique can be
extrapolated.  A related issue arises in the SQM description of D0-branes in the
presence of pure D8-branes, where it is necessary to add an additional 
D8-brane source to cancel IR divergences \refs{\dkps,\hetlife}.
Another possibility comes from considering the very recent 
work of \refs{\bdg, \dfk} where it was shown that when a D0-brane
moves through a D8-brane a fundamental fermionic 0-8 string is
created.\foot{In \refs{\dfk} it was argued that presence of ``half'' 
a fundamental string accounts for the attractive force coming from the
$(-1)^F$R open string sector, which is dual to the exchange of
R-R closed strings between the branes.} 
Accounting for this phenomena in M(atrix) theory may
resolve the discrepancy.
 
Following \refs{\power,\dealwis,\tasi} we can make some comments about the
consequences of interpreting the IIA D8-brane as a wrapped 11D
nine-brane.  If we compactify the 11D theory on a circle of radius
$R_{11} = \sqrt{\alpha'} g_s$, then upon double dimensional 
reduction the D8-brane tension is related to the tension of the 11D nine-brane by
(in units where $g_{s\mu\nu}=g_{11\mu\nu}$),

\eqn\Treduce{T_8 = {2 \pi \sqrt{\alpha'} g_s T_9^M}.}
The tension of a Dp-brane is also given in terms the string tension 
$T_s = {1\over 2\pi\alpha'}$ by the formula,

\eqn\TDbrane{T_p^2 = {1 \over g_s} (2 \pi)^{1-p} T_s^{p+1}.}
Equating \Treduce, \TDbrane and using the fundamental membrane tension
$2\pi R_{11} T_2^M = T_s$ leads to a relation between
11D quantities,

\eqn\Televen{T_9^M = {R_{11}^2 (T_2^M)^4 \over 2\pi}.}

In the $R_{11} \rightarrow \infty$ limit, $T_9^M$ diverges.  This can
be compared to an analogous calculation, treating the D6-brane as
an unwrapped 11D six-brane which leads to the tension formula,
$T_6 = T_6^M$.  Again, using \TDbrane and
trading $g_s$ and $\alpha'$ for $R_{11}$ and $T_2^M$, we can derive an
11D relation,

\eqn\Tsix{T_6^M = R_{11}^2 (T_2^M)^3.}
$T_6^M$ also diverges in the $R_{11} \rightarrow \infty$ limit, which is
consistent with its interpretation as a magnetic KK p-brane \refs{\townsend}.
The fact that $T_9^M$ behaves in the same way suggests that the 10D 8-brane
may have a similar 11D origin.\foot{The fact that the two may be related 
was previously mentioned in the context of massive 
IIA supergravity \refs{\dualbrane}.}  The M(atrix) six-brane and eight-brane
configurations were also shown to have energy densities which scale to 
infinity in the large $N$ limit in \refs{\mbranes}  
\foot{This is in contrast
to the ``end of the world" 9-branes which are related to an orientifold
plane plus 8 D8-branes and their mirror images and which should have
finite tensions in the uncompactified limit.}.


\bigskip\bigskip\centerline{{\bf Note Added}}\nobreak
After this work had already appeared we received \refs{\bgl}
which studies interactions of the pure D0-D8 system in
some detail.

\bigskip\bigskip\centerline{{\bf Acknowledgements}}\nobreak

I am grateful to Steve Giddings, Eric Gimon, Karl Landsteiner, Philippe Pouliot
and especially to Joe Polchinski for useful discussions.  
This work was supported in part by
DOE grant DOE-91ER40618 and
by NSF PYI grant PHY-9157463.

\listrefs

\end